\documentclass[12pt,notitlepage,a4paper]{article}

\usepackage{epsfig}
\usepackage{color,graphicx}
\usepackage{cite}
\usepackage{amssymb,amsmath}

\newcommand{\be}{\begin{equation}}
\newcommand{\ee}{\end{equation}}
\newcommand{\bea}{\begin{eqnarray}}
\newcommand{\eea}{\end{eqnarray}}

\begin{document}

\begin{flushright}
CCTP-2010-24 
\end{flushright}

\begin{center}  

\vskip 2cm 

\centerline{\Large {\bf Magneto-roton excitation}} 
\vskip 0.5cm 
\centerline{\Large{\bf in a holographic quantum Hall fluid}}
\vskip 1cm

\renewcommand{\thefootnote}{\fnsymbol{footnote}}

\centerline{Niko Jokela,${}^{1,2}$\footnote{najokela@physics.technion.ac.il} 
Gilad Lifschytz,${}^2$\footnote{giladl@research.haifa.ac.il} 
and Matthew Lippert${}^3$\footnote{mlippert@physics.uoc.gr}}

\vskip .5cm
${}^1${\small \sl Department of Physics} \\
{\small \sl Technion, Haifa 32000, Israel} 

\vskip 0.2cm
${}^2${\small \sl Department of Mathematics and Physics} \\
{\small \sl University of Haifa at Oranim, Tivon 36006, Israel} 

\vskip 0.2cm
${}^3${\small \sl Crete Center for Theoretical Physics} \\
{\small \sl Department of Physics} \\
{\small \sl University of Crete,  71003 Heraklion, Greece}

\end{center}

\vskip 0.3 cm

\setcounter{footnote}{0}
\renewcommand{\thefootnote}{\arabic{footnote}}

\begin{abstract}
\noindent  We compute the neutral bosonic excitation spectra of the holographic quantum Hall fluid described by the D3-D7' system. 
We find that the system is stable, gapped, and, in a range of parameters, exhibits a spectrum of low-lying excitations
very similar to that of a fractional quantum Hall fluid, including a magneto-roton excitation.

\end{abstract}

\newpage

\tableofcontents

\section{Introduction}
Strongly interacting fermions in $2+1$ dimensions exhibit many interesting phenomena which are not easily accessible with conventional theoretical 
tools.  Gauge/gravity duality offers a new tool which may be better suited for describing and maybe advancing our understanding of these phenomena.  
In \cite{Bergman:2010gm}, a holographic system exhibiting a mass gap at nonzero density and magnetic field was described. 
This system appears to have many of the properties of a fractional quantum Hall fluid.  Other holographic models of the quantum Hall effect 
can be found in \cite{otherqh}. 

The fluctuation spectra of quantum Hall fluids have many interesting properties, such as gaps for both charged and 
neutral excitations and fractional charged excitations with anyonic statistics.   One interesting feature, which has not received much attention 
from holographic studies, is that the lowest energy neutral excitation is a magneto-roton.  
Magneto-rotons are collective excitations whose dispersion 
relations have a minimum at some nonzero momentum.  Although at this minimum the magneto-rotons carry momentum, their group velocity vanishes.  
Experimentally, these are seen in inelastic light scattering experiment \cite{light} and 
ballistic phonons \cite{ballistic}, and most recently in \cite{recent}. Theoretically, such magneto-roton excitations have been analysed 
using an analogy with the Feynman-Bijl wave-function for the collective excitation of the 
superfluid in \cite{Girvin:1986zz},\footnote{Terminology for the roton is borrowed from studies of the excitations of superfluids, where the phonon
dispersion has a local minimum at a high momentum.} the 
formulation in terms of composite fermions in \cite{spj}, as a Chern-Simons effective theory in \cite{cs}.

In this paper we study the neutral bosonic excitation spectra of the D3-D7' system described in \cite{Bergman:2010gm}. 
We find that the system is stable and gapped. We also find that the lowest energy excitation in a certain range of densities 
displays a magneto-roton minimum at a finite momentum. 


The sectioning of the paper is the following. In Section \ref{sec:setup} we present the model of \cite{Bergman:2010gm} in a set of coordinates more suitable for 
fluctuation analysis. In Section \ref{sec:fluct} we lay out the equations of motion of the fluctuations and the method for solving them. 
In Section \ref{sec:numerics} we present the results. We conclude with a few comments.

\section{Review of the model}\label{sec:setup}

In \cite{Bergman:2010gm} a holographic dual to a (2+1)-dimensional system of strongly-coupled fermions in the fundamental representation was 
proposed. The brane configuration in \cite{Bergman:2010gm} consists of a probe D7-brane in the background of $N\to \infty$ D3-branes.
The intersection of the D7-brane with the D3-branes is (2+1)-dimensional, such that $\#NN=3$ and $\#ND=6$, and hence the massless spectrum of the D3-D7 open strings 
consists only of the desired fermions (see \cite{Rey:2008zz} for earlier consideration of this model, and for its relationship to 
three-dimensional QCD, see \cite{Hong:2010sb}).  One can then give the fermions mass by separating the branes in the transverse direction.  

We decouple these fermions from the closed string modes by focusing on the near-horizon limit of a large stack of $N$ D3-branes and treating 
the D7-brane in the probe limit.  According to the usual gauge/gravity duality, the dynamics of the probe D7-brane in the near-horizon D3-brane 
background captures the physics of these (2+1)-dimensional fermions interacting with a strongly-coupled (3+1)-dimensional gauge theory. 
We will review the details of this D3-D7' model and how to obtain stable Minkowski (MN) embeddings, which do not enter the horizon, of the 
D7-brane probe at zero temperature.

\subsection{Action}

The D3-brane background has the metric
\be 
 ds_{10}^2 = \frac{r^2}{L^2} \left(-h(r)dt^2+dx^2+dy^2+dz^2 \right)+\frac{L^2}{r^2}\left(\frac{dr^2}{h(r)}+r^2 d\Omega_5^2\right) \ ,
\ee 
where $h(r)=1-\frac{r_T^4}{r^4}$ and where we parameterize the $S^5$ as
\be
 d\Omega_5^2 = d\psi^2 + \cos^2\psi(d\theta_1^2+\sin^2\theta_1 d\phi_1^2)+\sin^2\psi(d\theta_2^2+\sin^2\theta_2 d\phi_2^2) \ .
\ee
The AdS radius is given by $L^2=\sqrt{4\pi g_{s} N}\alpha'=\sqrt{\lambda}\alpha'$, and the temperature is $T = \frac{r_T}{\pi L^2}$.
In addition, the RR four-form $C^{(4)}_{txyz}=-\frac{r^4}{L^4}$.  The angles have the following 
ranges: $\psi \in [0, \pi/2]$, $\theta_{1,2} \in [0, \pi]$, and $\phi_{1,2} \in [0, 2\pi]$.
In the following, we will only be interested in zero temperature, so we set
\be
 r_T = 0 \ .
\ee

With a view towards the fluctuation analysis, it is useful to introduce the following Cartesian coordinates:
\bea
 \rho & = & r \sin\psi \\
 R    & = & r \cos\psi \ .
\eea
We take the D7-brane probe to be wrapped on the two $S^2$'s and to span the $t,x,y$, and $R$ directions and assume that the embedding 
fields $z$ and $\rho$ are only functions of $R$. In order to have a stable configuration, we need to turn on a magnetic flux through one of 
the $S^2$ \cite{Myers:2008me,Bergman:2010gm}; we choose
\be
2\pi\alpha' F_{\theta_2\phi_2} = \frac{f L^2}{2}\sin\theta_2 \ .
\ee

It turns out that with flux only on one of the $S^2$'s a constant $z$ is a solution to the equations of motion; we can thus set it to zero in what follows. 
The induced metric on the D7-brane is 
\bea
 ds^2_{D7} = \frac{r^2}{L^2} \left(-dt^2+dx^2+dy^2\right)+\frac{L^2}{r^2} \left(1+\rho'^2\right)dR^2+\frac{L^2R^2}{r^2}d\Omega_2^{(1)2}+\frac{L^2\rho^2}{r^2}d\Omega_2^{(2)2} \ .
\eea
Since we are interested in obtaining a quantum Hall description, we will turn on a background charge density and a magnetic field:
\be
 2\pi\alpha' F_{xy} = b \ \ , \ \ 2\pi\alpha' F_{Rt} = a'_0 \ . 
 \ee
The D7-brane DBI action reads:
\be
 S = -\tilde{\mathcal N}\int dR \ R^2\sqrt{f^2+4\frac{\rho^4}{r^4}}\sqrt{1+\frac{b^2L^4}{r^4}}\sqrt{1+\rho'^2-a'^2_0} \ ,
 \ee
where $\tilde{\mathcal N} = 8\pi^2 T_7 V_3 L^2$. The CS action is \cite{Bergman:2010gm}
\be
 S_{CS} = +\tilde{\mathcal N} L^2  \int dR \ b a'_0  c(R) \ ,
\ee
where the pull-back of the RR-potential onto the worldvolume reads
\be
 c(R) = \arctan\left(\frac{\rho}{R}\right)-\frac{1}{4}\sin\left(4\arctan\left(\frac{\rho}{R}\right)\right)-\psi_\infty+\frac{1}{4}\sin(4\psi_\infty) \ .
\ee
The asymptotic angle $\psi_\infty =\lim_{R \rightarrow \infty}\arctan(\frac{\rho}{R})$ is related to $f$ by
\be
 f^2=4\sin^{2} \psi_{\infty}-8\sin^{4} \psi_{\infty} \ .
\ee

A rescaling of the coordinates and the temporal component of the gauge field
\be\label{eq:rescaling}
 R = L\sqrt b \tilde R \ , \ \rho = L\sqrt b \tilde\rho \ , \ r = L\sqrt b \tilde r\ , \ a_0 = L\sqrt b \tilde a_0 \ ,
\ee
enables us to write down the full D7-brane action in a `magnetic-fieldless' fashion
\be
 S = -\mathcal N \int d\tilde R \left( \tilde R^2\sqrt{f^2+4\frac{\tilde\rho^4}{\tilde r^4}}\sqrt{1+\frac{1}{\tilde r^4}}\sqrt{1+\tilde\rho'^2-\tilde a'^2_0} + \tilde a'_0 c(\tilde R) \right) \ , 
\ee
where $\mathcal N \equiv b^{3/2}L^3\tilde{\mathcal N}$ and the prime now indicates derivative with respect to $\tilde R$.

\subsection{Equations of motion}
The $\tilde a'_0$ equation of motion, integrated once, reads:
\be
\label{a0eom}
 \tilde g \left(1+\frac{1}{\tilde r^4}\right)\tilde a'_0  = \frac{d}{b}-2 c(\tilde R) \equiv \frac{\tilde d}{b} \ ,
\ee
where
\be
\label{gdef}
 \tilde g = \frac{2\tilde R^2\sqrt{f^2+4\frac{\tilde \rho^4}{\tilde r^4}}}{\sqrt{1+\tilde r^{-4}}\sqrt{1+\tilde \rho'^2-\tilde a'^2_0}} \ .
\ee
The constant of integration $d$ was chosen to match the notation of \cite{Bergman:2010gm} and is related to the physical charge 
density by $D = (2\pi\alpha'/L) (\mathcal N/V_3) d$.  After a little bit of an algebra, equations (\ref{a0eom}) and (\ref{gdef}) can be put in the form
\bea
 \tilde g    & = & \frac{1}{1+\tilde r^{-4}}\sqrt{\frac{\left(\frac{\tilde d}{b}\right)^2+4\tilde R^4(1+\tilde r^{-4})(f^2+4\frac{\tilde\rho^4}{\tilde r^4})}{1+\tilde\rho'^2}} \\
 \tilde a'_0 & = & \frac{\tilde d}{b}\sqrt{\frac{1+\tilde\rho'^2}{\left(\frac{\tilde d}{b}\right)^2+4\tilde R^4(1+\tilde r^{-4})(f^2+4\frac{\tilde\rho^4}{\tilde r^4})}} \ .
\eea
The $\tilde\rho$ equation of motion is:
\be\label{eq:rhoEOM}
 \partial_{\tilde R}\left(\tilde g \left(1+\frac{1}{\tilde r^4}\right)\tilde\rho'\right) = -16\frac{\tilde R^3\tilde\rho^2 \tilde a'_0}{\tilde r^6}+\frac{32\tilde R^6\tilde\rho^3}{\tilde g\tilde r^6}-\frac{8\tilde R^4\tilde \rho \left(f^2+4\frac{\tilde\rho^4}{\tilde r^4}\right)}{ \tilde g \tilde r^6\left(1+\frac{1}{\tilde r^4}\right)} \ .
\ee

We are interested in Minkowski embeddings, for which the D7-brane does not enter the horizon. 
These are characterized by the IR boundary condition $\tilde\rho(\tilde R=0) = \tilde\rho_0$. 
We solve the $\tilde\rho$ equation of motion numerically, shooting from the tip $\tilde R_0=0$, using a smooth 
boundary condition $\tilde\rho'(0)=0$, toward the UV boundary.  As shown in \cite{Bergman:2010gm}, 
the ``fermion mass" $m$ and ``chiral condensate" $c$ are holographically encoded in the UV 
expansion of $\psi$:\footnote{The $m$ and $c$ are not really mass and condensate for $\Delta_+ \ne -1$.}
\be
\psi = \psi_\infty + m r^{\Delta_+} - c r^{\Delta_-}+ \ldots \ , 
\ee
where
\be
\label{Deltapm}
 \Delta_\pm=-\frac{3}{2}\pm \frac{1}{2}\sqrt{73-\frac{48}{\cos^2\psi_\infty}} \ .
\ee
In the rescaled Cartesian coordinates used here, the rescaled mass, $\tilde m = m/(L\sqrt b)$, can be obtained from the numerically-obtained solution for $\tilde \rho(\tilde R)$: 
\be
\tilde m = \lim_{\tilde R\to\infty} \tilde r^{-\Delta_{+}}\sin\left(\arctan\left(\frac{\tilde\rho}{\tilde R}\right)-\psi_\infty\right) \ .
\ee

As explained in \cite{Bergman:2010gm}, for Minkowski embeddings, the charge density $d$ and the magnetic field $b$ are not independent but locked:
\be
 \frac{d}{b} = 2 c(0) = \frac{\pi\nu}{N} \ ,
\ee
where $\nu/N$ is the filling fraction per fermion species, which is  
fixed by the internal flux $f$. The quantization of the filling  
fraction is then a direct consequence of the quantization of the  
wrapped flux.  The equation of motion for $\tilde\rho$ (\ref{eq:rhoEOM}) fixes $\psi_\infty
$ in terms of $f$, which, through equation (\ref{Deltapm}), also fixes $\Delta_+
$.  Recall from \cite{Bergman:2010gm} that the filling fraction is 
restricted to lie in the range
\be\label{eq:range}
 0.6972 \lesssim \frac{\nu}{N}\lesssim 0.8045
\ee
and does not take the usual rational values considered in experimental setups.
Particular values of the filling fraction for which we will present numerical results are:
\bea
 \Delta_+ & = -\frac{1}{2} \ : \qquad & \frac{\nu}{N} \approx 0.7426 \\
 \Delta_+ & = -1 \ : \qquad & \frac{\nu}{N} \approx 0.7082 \\
 \Delta_+ & = -\frac{5}{4} \ : \qquad & \frac{\nu}{N} \approx 0.6999 \ .
\eea

\begin{figure}[ht]
\centerline{\epsfig{file=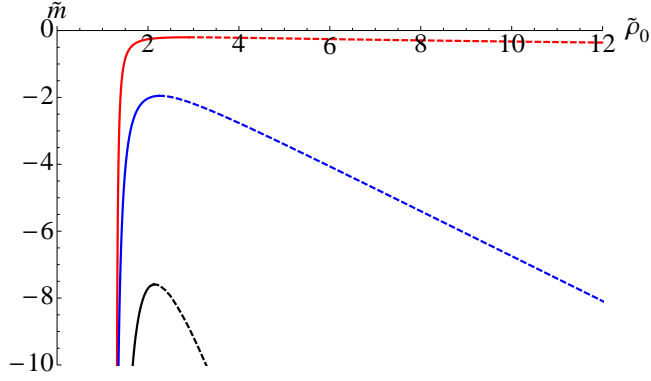,height=5cm}}
\caption{$\tilde m$ vs $\tilde\rho_0$ for MN embeddings. The solid line is the stable MN embedding, and the dashed line is the unstable MN 
embedding. The top, red curve represents $\Delta_{+}=-1/2$, the middle, blue curve $\Delta_{+}=-1$, and the bottom, black curve $\Delta_{+}=-5/4$.}
\label{fig:mvsrho0}
\end{figure}

The dependence of $\tilde m$ on $\tilde \rho_0$ is illustrated in Fig.~\ref{fig:mvsrho0} for 
three different choices of $\Delta_{+}$.\footnote{We expect there to be also zero-temperature ``black hole solutions" reaching 
the origin of AdS, taking any values of the mass \cite{Bergman:2010gm}.}  Note that for fixed (non-rescaled) mass $m$, lower 
values of $\tilde m$ correspond to smaller background magnetic fields $b$.  
For a given $\tilde m$, the stable Minkowski embedding is the one with lower $\tilde\rho_0$, as was shown by energetic 
considerations in \cite{Bergman:2010gm}. We will confirm this by showing that the larger $\tilde\rho_0$ Minkowski embedding has a tachyon 
in its spectrum, whereas for the smaller $\tilde\rho_0$ it is absent. At $b=0$ only the unstable Minkowski embedding exists.

\section{Fluctuations}\label{sec:fluct}

We wish to obtain the spectrum of small fluctuations around the classical 
solution $\tilde{\bar \rho}(\tilde R),\tilde{\bar a}_0(\tilde R)$, and $\tilde{\bar a}_y(x) = \frac{\sqrt b}{L} x$. The energy of the lowest normalizable fluctuation gives 
us the energy gap above the quantum Hall fluid. As before, we write the equations of motion and background solutions in the 
rescaled variables (\ref{eq:rescaling}) and rescale also the fluctuations:
\bea
 \delta a_\mu = \sqrt b L \delta \tilde a_\mu \\
 \delta\rho = \sqrt b L \delta\tilde\rho \ ,
\eea
where $\mu = t,x,y,\tilde R$. We need to allow for fluctuations of (almost) all worldvolume fields:
\bea
 z           & = & \epsilon\delta z(t,x,y,\tilde R) \\
 \tilde\rho        & = & \tilde{\bar\rho}(\tilde R)+\epsilon\delta\tilde \rho(t,x,y,\tilde R) \\
 \tilde a_t         & = & \tilde{\bar a}_0(\tilde R)+\epsilon\delta\tilde a_t(t,x,y,\tilde R) \\
 \tilde a_x         & = & \epsilon\delta\tilde a_x(t,x,y,\tilde R) \\
 \tilde a_y         & = & \frac{\sqrt b}{L}  x +\epsilon\delta\tilde a_y(t,x,y,\tilde R) \\
 \tilde a_{\tilde R}& = & \epsilon\delta\tilde a_{\tilde R}(t,x,y,\tilde R) \ ,
\eea
with $\epsilon$ a small parameter.  We do not consider fluctuations in the internal spheres or dependence on the internal two two-sphere 
coordinates. From now on, we will omit the bars from the background fields.

We expand the Lagrangian to second order in $\epsilon$:
\be
 \mathcal L = \mathcal L_0 + \epsilon\mathcal L_1 + \epsilon^2\mathcal L_2 + \ldots \ .
\ee
Here $\mathcal L_0$ corresponds to the background on-shell Lagrangian and $\mathcal L_1$ vanishes by the background equations of motion. 
From $\mathcal L_2$ one can derive the equations of motion for fluctuations. 

Not all of the above fluctuations are physical, however. Since we are working in the gauge $\tilde a_{\tilde R} = 0$, we must 
demand that $\delta\tilde a_{\tilde R} = 0$, too.
The equation of motion for $\delta\tilde a_{\tilde R}$, after imposing $\delta \tilde a_{\tilde R} = 0$, gives us a constraint 
which the other fluctuations have to obey.

We are interested in wavelike solutions for the fluctuation spectrum:
\bea
 \delta z & \sim & e^{-i\omega^{(z)} t+ik^{(z)}_x x+ik^{(z)}_y y} \\
 \delta \tilde a_\mu  & \sim & e^{-i\omega t+ik_x x+ik_y y} \\
 \delta \tilde\rho & \sim & e^{-i\omega t+ik_x x+ik_y y} \ ,
\eea
where, anticipating the decoupling of $\delta z$ fluctuation from the rest, we introduce a separate frequency $\omega^{(z)}$ and 
two-momentum $k^{(z)}$ from the others. All the other fluctuations couple to one another at nonzero momentum.  
Rescaling the three-momenta as well,
\bea
  \tilde \omega^{(z)} = \frac{L}{\sqrt b}\omega^{(z)} &, &  \tilde k^{(z)}_{x,y} = \frac{L}{\sqrt b}k^{(z)}_{x,y} \\
  \tilde \omega = \frac{L}{\sqrt b} \omega \ \ \ &,& \tilde k_{x,y} = \frac{L}{\sqrt b}k_{x,y} \ ,
\eea
makes the resulting equations of motion dimensionless.

We also make use of the rotation symmetry in the $(x,y)$-plane and choose the vectors $\vec k^{(z)}$ and $\vec k$ to be aligned along $x$-axis. From now on, we
omit the subscript $x$ from the momenta.

\subsection{Equations of motion}\label{sec:EOM}

To make sure that all the fluctuations are physical, we write the equations of motion in a manifestly gauge invariant way. That is, we introduce 
a gauge invariant combination: $\delta\tilde e_x = (\tilde k\delta\tilde a_t+\tilde\omega\delta \tilde a_x)$.
The equations of motion for $\delta\tilde a_t$ and $\delta\tilde e_x$ are
\bea
 \partial_{\tilde R}\tilde H &  = & -i\tilde k\delta\tilde a_y\partial_{\tilde R}\left(\tilde g\tilde a'_0(1+2\tilde r^{-4})\right)-\frac{\tilde g}{\tilde r^4}(1+\tilde\rho'^2)\tilde k\delta\tilde e_x+\frac{\tilde g}{\tilde r^4}\tilde a'_0\tilde\rho'\tilde k^2\delta\tilde\rho   \nonumber\\
 \partial_{\tilde R}\left(\frac{\tilde g}{\tilde\omega} (\delta\tilde e'_x - \tilde k\delta \tilde a'_t)\right) 
  & = & -i\tilde\omega \delta\tilde a_y\partial_{\tilde R}\left(\tilde g\tilde a'_0(1+2\tilde r^{-4})\right)-\frac{\tilde g}{\tilde r^4}(1+\tilde\rho'^2)\tilde\omega\delta\tilde e_x +\frac{\tilde g}{\tilde r^4}\tilde a'_0\tilde\rho'\tilde \omega\tilde k\delta\tilde\rho \nonumber\ .
\eea
The equation for $\delta\tilde a_y$ is
\bea
i\delta\tilde e_x\partial_{\tilde R}\left(\tilde g\tilde a'_0(1+2\tilde r^{-4})\right)-\tilde\omega^2\frac{\tilde g}{\tilde r^4}(1+\tilde\rho'^2)\delta\tilde a_y -\partial_{\tilde R}\left(\tilde g \delta\tilde a'_y\right)+\tilde k^2\frac{4\tilde R^4\left(f^2+4\frac{\tilde\rho^4}{\tilde r^4}\right)}{\tilde g \tilde r^4\left(1+\tilde r^{-4}\right)^2}\delta\tilde a_y \nonumber\\
 =  i\tilde k\delta\tilde\rho\partial_{\tilde R}\left(\frac{\tilde g }{\tilde r^4}\tilde\rho' \right)+i\tilde k\frac{16\tilde R^3\tilde\rho^2}{\tilde r^6}\tilde a'_0 \delta\tilde\rho-i\tilde k\tilde N \delta\tilde\rho  \ ,
\eea
and for $\delta\tilde\rho$ we have
\bea
 &   & \partial_{\tilde R}\left[\frac{\tilde g}{\tilde A}(1-\tilde a'^2_0)\left(1+\tilde r^{-4}\right)\delta\tilde\rho'\right]+\left\{\tilde M -\partial_{\tilde R}[\tilde K\tilde\rho']-\frac{32\tilde a'_0\tilde\rho \tilde R^3}{\tilde r^8}\left(2\tilde\rho^2-\tilde R^2\right)\right\}\delta\tilde\rho \nonumber\\
 &   & +\left\{\tilde\omega^2\frac{\tilde g}{\tilde r^4}\left(1+\tilde r^{-4}\right)-\tilde k^2\frac{\tilde g}{\tilde r^4}\left(1-\tilde a'^2_{0}\right) \right\}\delta\tilde\rho \nonumber\\
 & = & \left(\tilde K\tilde a'_0+\frac{16\tilde R^3}{\tilde r^6}\tilde\rho^2\right)\delta\tilde a'_t-\partial_{\tilde R}\left[\frac{\tilde g}{\tilde A}\left(1+\tilde r^{-4}\right)\tilde a'_0\tilde\rho'\delta\tilde a'_t\right] \nonumber\\
 &   & + \frac{\tilde g}{\tilde r^4}\tilde a'_0\tilde\rho'\tilde k\delta\tilde e_x +i\tilde k\left(-\tilde N+\frac{16\tilde R^3}{\tilde r^6}\tilde\rho^2\tilde a'_0+\partial_{\tilde R}\left[\frac{\tilde g}{\tilde r^4}\tilde\rho'\right]\right)\delta\tilde a_y \ .
\eea
As mentioned previously, the $\delta z$ equation of motion decouples from all the rest:
\be
 \partial_{\tilde R}\left[\tilde g\left(\tilde r^4+1\right)\delta z'\right] = -\tilde g\left(1+\tilde r^{-4}\right)\left(1+\tilde\rho'^2\right)\tilde\omega^{(z)2}\delta z +\tilde g\tilde A\tilde k^{(z)2}\delta z \ .
\ee
Finally, the constraint reads
\be\label{eq:constraint}
 \frac{\tilde k}{\tilde\omega}\tilde g\left(\delta\tilde e'_x-\tilde k\delta\tilde a'_t\right)-\tilde\omega\tilde H = 0 \ .
\ee
In the above equations, we have introduced the following functions which depend only on the background:
\bea
 \tilde A & = & 1+\tilde\rho'^2-\tilde a'^2_0 \\
 \tilde K & = & \frac{2\tilde g\tilde\rho}{\tilde r^{10}\left(f^2+4\frac{\tilde\rho^4}{\tilde r^4}\right)}\left(-4\tilde r^2 \tilde \rho^2-4\tilde r^6 \tilde \rho^2+8\tilde \rho^4+\tilde r^4(f^2+4\tilde \rho^4) \right) \\
 \tilde N & = & -\frac{8 \tilde R^4}{\tilde g \tilde r^{14}\left(1+\tilde r^{-4}\right)^2}\left(f^2\tilde r^4(1+2\tilde r^4)-4\tilde r^2 \tilde \rho^2(1+\tilde r^{4})+4\tilde\rho^4(2+3\tilde r^4) \right)\tilde\rho \\
 \tilde M & = & \frac{8\tilde R^4}{\tilde g\tilde r^{20}\left(1+\tilde r^{-4}\right)^2\left(f^2+4\frac{\tilde\rho^4}{\tilde r^4}\right)} \Bigg(f^4 \tilde r^{10}(1+\tilde r^4)-2f^2\tilde r^8 \tilde \rho^2(6+2f^2+3\tilde r^4 (4+f^2) +6\tilde r^8) \nonumber\\
 & & +12f^2\tilde r^6 \tilde \rho^4 (5+8\tilde r^4 +3 \tilde r^8)-8\tilde r^4 \tilde \rho^6(2+9f^2+2\tilde r^4(2+7f^2)+\tilde r^8 (2+3f^2)) \nonumber\\
 & & +80\tilde r^2 \tilde \rho^8(2+3\tilde r^4 +\tilde r^8)-32\tilde \rho^{10}(6+9\tilde r^4 +2\tilde r^8)\Bigg) \ .
\eea
In addition, the fluctuation-dependent function $\tilde H$ is defined by
\bea
 \tilde H & = & -\frac{\tilde g}{\tilde A}(1+\tilde r^{-4})(1+\tilde\rho'^2)\delta\tilde a'_t+\frac{\tilde g}{\tilde A}(1+\tilde r^{-4})\tilde a'_0\tilde\rho'\delta\tilde\rho'\nonumber\\
 & & +16\frac{\tilde R^3\tilde\rho^2}{\tilde r^6}\delta\tilde\rho+\tilde K \tilde a'_{0}\delta\tilde\rho \ .
\eea

Looking at the above equations, we notice that after using the constraint, the 
equations of motion are not independent. Indeed, one can see that using the constraint to eliminate $\delta\tilde a'_t$, the equations of motion 
for $\delta\tilde a_t$ and $\delta\tilde e_x$ become identical. So, in fact, we are left with only three equations for three 
unknown functions. Alternatively, one can solve four equations for all four functions and 
impose the constraint only on the boundary conditions.
 

\subsection{Methodology}\label{sec:method}
We are seeking those solutions which represent normalizable modes. 
In this parametrization, normalizability is translated to the requirement that at the boundary, as $\tilde R \to \infty$,
\begin{equation}
\delta \tilde a_y \rightarrow 0 \ \ , \ \  \delta\tilde e_x \rightarrow 0 \ \ ,\ \ 
\tilde R^{-1-\Delta_+}\delta\tilde\rho \rightarrow  0 \ .
\end{equation}
At the tip, the physical boundary conditions are that all derivatives with respect to $\tilde R$ vanish:
\be\label{eq:physical}
 \delta\tilde\rho'(0) = \delta\tilde e'_x(0) = \delta\tilde a'_y(0) = 0 \ .
\ee
The vanishing of $\delta\tilde\rho'(0)$ means that the embedding is smooth, and the boundary conditions for the gauge fluctuations ensure that there are no sources at the tip. 

The equations of motion (other than for $\delta z$) cannot be completely decoupled,\footnote{At $\tilde k=0$, $\delta\tilde \rho$  
decouples from the rest.} which makes our goal of finding solutions which obey normalizable boundary conditions at infinity much more difficult.  Operationally, we solve the equations by choosing boundary conditions at the tip and shooting out to infinity.  To find the desired normalizable solutions, we must systematically vary the choice of tip boundary conditions until we find those for which all the coupled fluctuations will simultaneously vanish at infinity. 

To accomplish this task, one can utilize a known determinant method \cite{Amado:2009ts, Kaminski:2009dh}.
We have specified three boundary conditions at the tip for the derivatives of the fluctuations. Since, we have three independent functions to solve for, and they 
satisfy ordinary, linear, second-order homogeneous differential equations, we should choose three linearly-independent boundary conditions for the 
fluctuations themselves. Following the method of \cite{Amado:2009ts}, we choose a set of linearly-independent boundary conditions, \emph{e.g.}, 
\be\label{eq:bound}
 (\delta\tilde\rho(0),\delta\tilde e_x(0),\delta\tilde a_y(0)) = \{(1,0,0),(0,1,1),(0,1,-1)\} \ ,
\ee
which span the space of all possible boundary conditions at the tip.  We then solve the equations of motion numerically for a fixed $(\tilde\omega,\tilde k)$ for each 
group and compute the following determinant:
\be
 \det\left( \begin{array}{ccc}
 \tilde R^{-\Delta_{+}-1}\delta\tilde \rho^I & \tilde R^{-\Delta_{+}-1}\delta\tilde \rho^{II} & \tilde R^{-\Delta_{+}-1}\delta\tilde \rho^{III} \\
 \delta\tilde e_x^I & \delta\tilde e_x^{II} & \delta\tilde e_x^{III} \\
 \delta\tilde a_y^I & \delta\tilde a_y^{II} & \delta\tilde a_y^{III} 
\end{array} \right)\Bigg|_{\tilde R\to \infty} \ ,
\ee
where the Roman index corresponds to the given group of initial conditions in (\ref{eq:bound}).  For a given $\tilde k$, we then scan through $\tilde\omega$ 
($\tilde\omega$ can also take complex values) and calculate the determinant at each step. When the determinant is zero, 
there is a linear combination of the initial conditions giving the desired normalizable solution for which all fluctuations vanish at infinity.  Said differently, if for the set $(\tilde \omega, \tilde k)$ there is such a solution, it means that at $(\tilde \omega,\tilde  k)$ there is a pole in the Green's functions \cite{Amado:2009ts}.

\section{Numerical results}\label{sec:numerics}
We first present the numerical analysis for the case $\tilde k=0$. For this case, in addition to the generic 
decoupling of $\delta z$, the $(\delta\tilde e_x,\delta\tilde a_y)$ system and the $(\delta\tilde\rho, \delta\tilde a_t)$ system 
decouple from each other as well.
The energy of the normalizable excitation represents the gap to the next energy state. 
At zero $\tilde k$, we can identify $(\delta\tilde e_x,\delta\tilde a_y)$ as the vector excitation and $(\delta\tilde\rho, \delta\tilde a_t)$ as the scalar excitation, 
but since they mix at nonzero $\tilde k$, such a distinction is not generically possible. Nevertheless, we can continue to refer 
to these modes by their $\tilde k=0$ labels. 

\begin{figure}[ht]
\center
\includegraphics[width=0.30\textwidth]{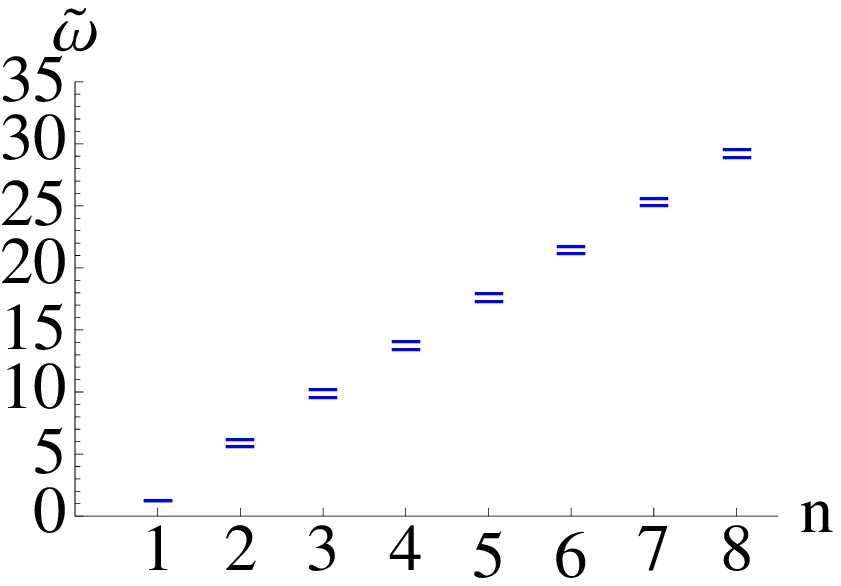}
\includegraphics[width=0.30\textwidth]{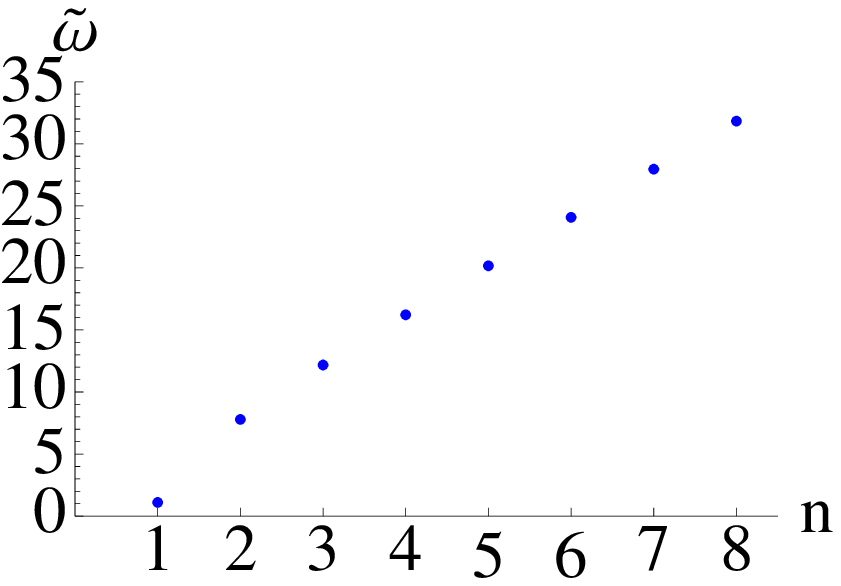}
\includegraphics[width=0.30\textwidth]{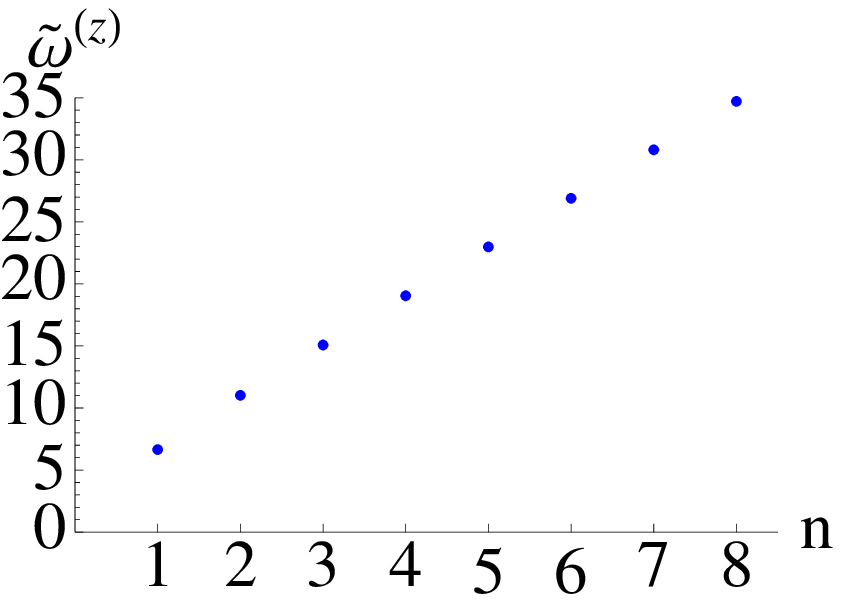}
\caption{The mass (energy gap) at $\tilde k=0$ in units of the magnetic length $\frac{\sqrt{b}}{L}$ for the first few excitations. 
In the left panel we show the excitations in the vector sector, middle panel shows the excitations in 
the scalar sector, and the right panel the excitations in the scalar $\delta z$. All figures are 
generated at $\Delta_+=-1$ and $\tilde m = -2.185$.}
\label{wfirsts}
\end{figure}

Fig.~\ref{wfirsts} shows the first few excitations of the various fields. First, we do not find any tachyons, implying that the embedding is stable. 
We also see that the embedding describes a gapped state in the neutral sector which means that it is an incompressible fluid. The spectrum of the excitations 
is approximately evenly spaced with a scale set by $\frac{\sqrt{B}}{(g_{YM}^{2}N)^{1/4}}$, where $B$ is the physical magnetic field.  Note 
that in the vector sector, the excitations come in doublets except for the lowest one.
 
\begin{figure}[ht]
\center
\includegraphics[width=0.45\textwidth]{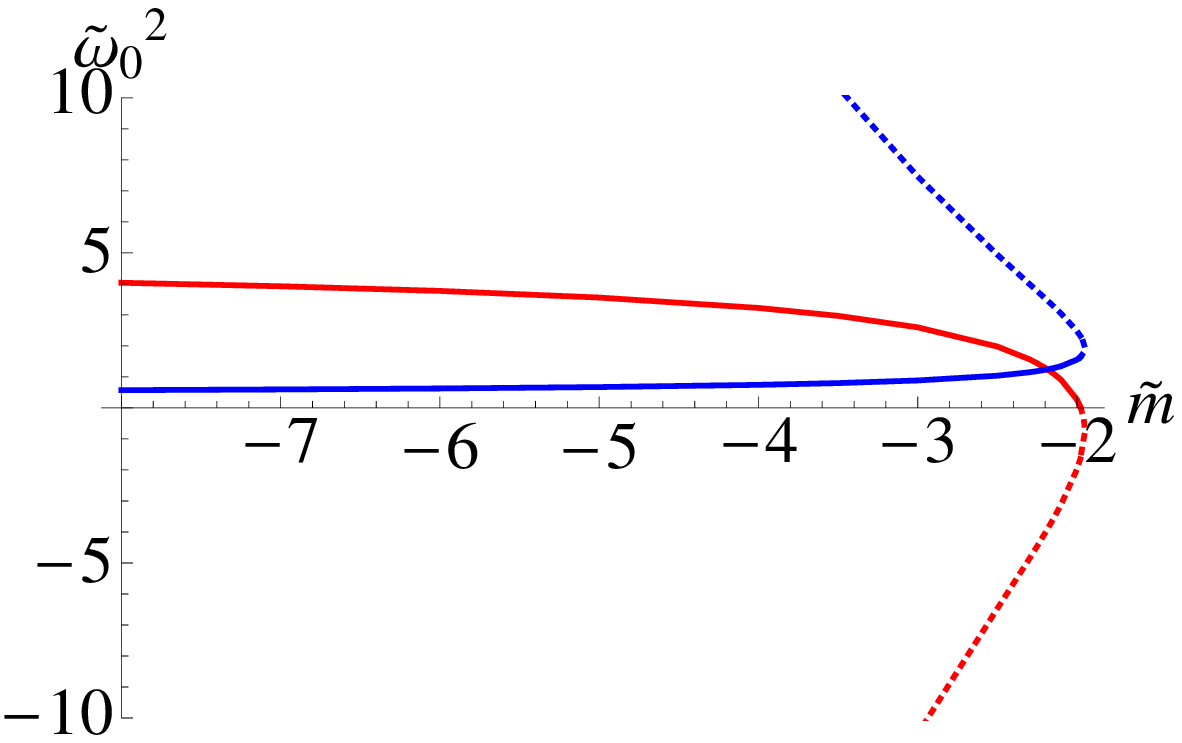}
\includegraphics[width=0.40\textwidth]{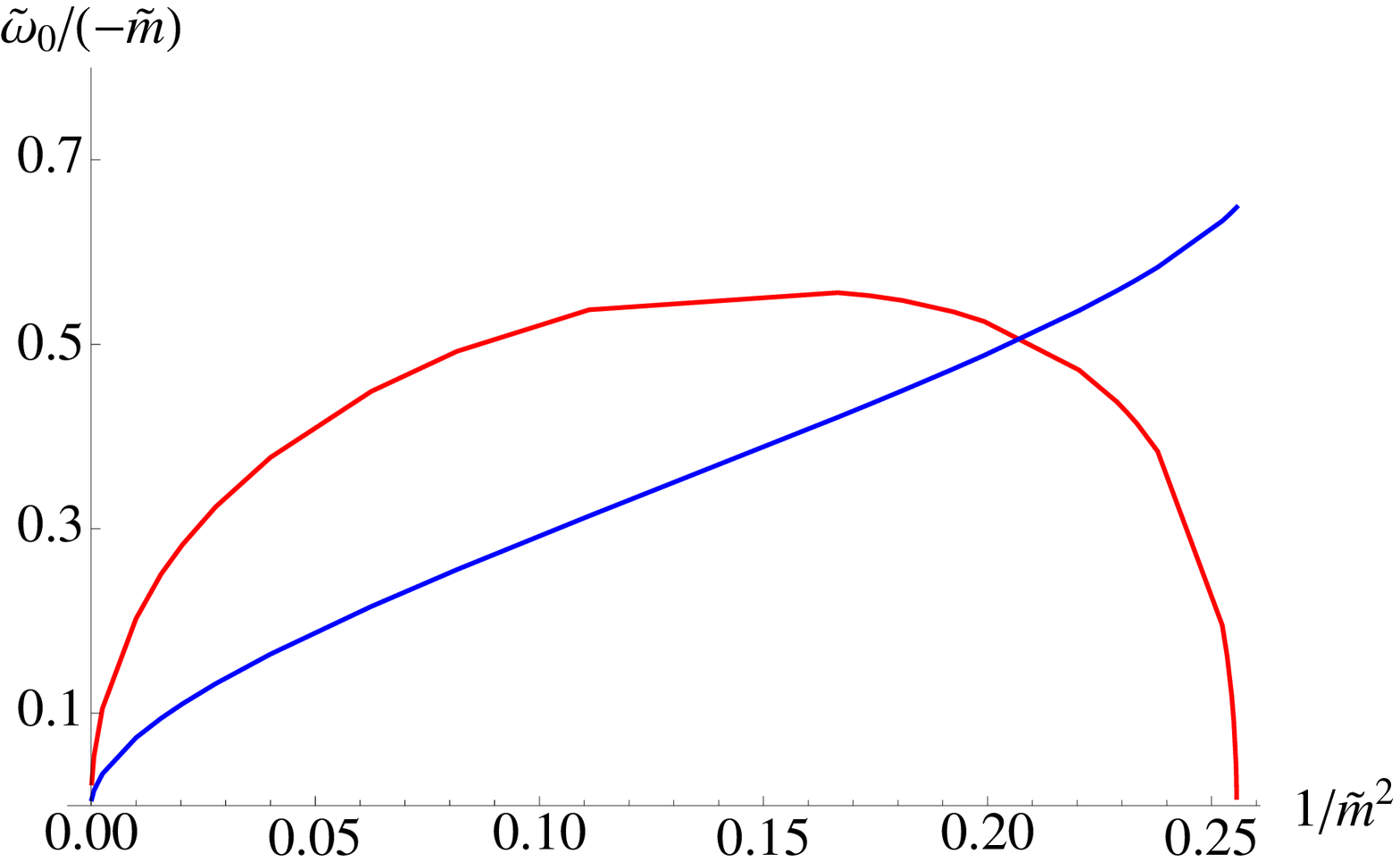}
\caption{The mass (energy gap) squared at $\tilde k=0$ in units of the magnetic length, of the two lowest energy excitations at zero momentum are depicted in the left panel. 
The red (blue) line represents the scalar (vector) fluctuation. Dashed lines are the same fluctuations in the unstable MN embedding. Notice that the scalar fluctuation
is tachyonic in the unstable MN branch. In the right panel, the energy gap at $\tilde k=0$ as a function of the magnetic field (or density). The vector fluctuation
increases in the whole range of the magnetic field. Both figures are generated at $\Delta_+=-1$.}
\label{wlowest}
\end{figure}

For most excitations, $\tilde\omega$ grows when the magnetic field increases (at fixed filling fraction).  However, the energy of the lowest scalar 
excitation decreases at large enough magnetic field. 
To illustrate this, we plot in Fig.~\ref{wlowest} the energy, in units of the magnetic length $\frac{\sqrt{b}}{L}$, of the lowest scalar mode and 
the lowest vector mode. At small magnetic field, the vector has lower mass, but as the magnetic field grows, the scalar mass 
drops down and eventually vanishes.  This occurs for large magnetic fields close to where the branch of the stable MN solution 
meets the branch of the unstable solution. Going onto the unstable branch, we see the tachyon explicitly as the scalar embedding fluctuation.

\begin{figure}[ht]
\center
\includegraphics[width=0.45\textwidth]{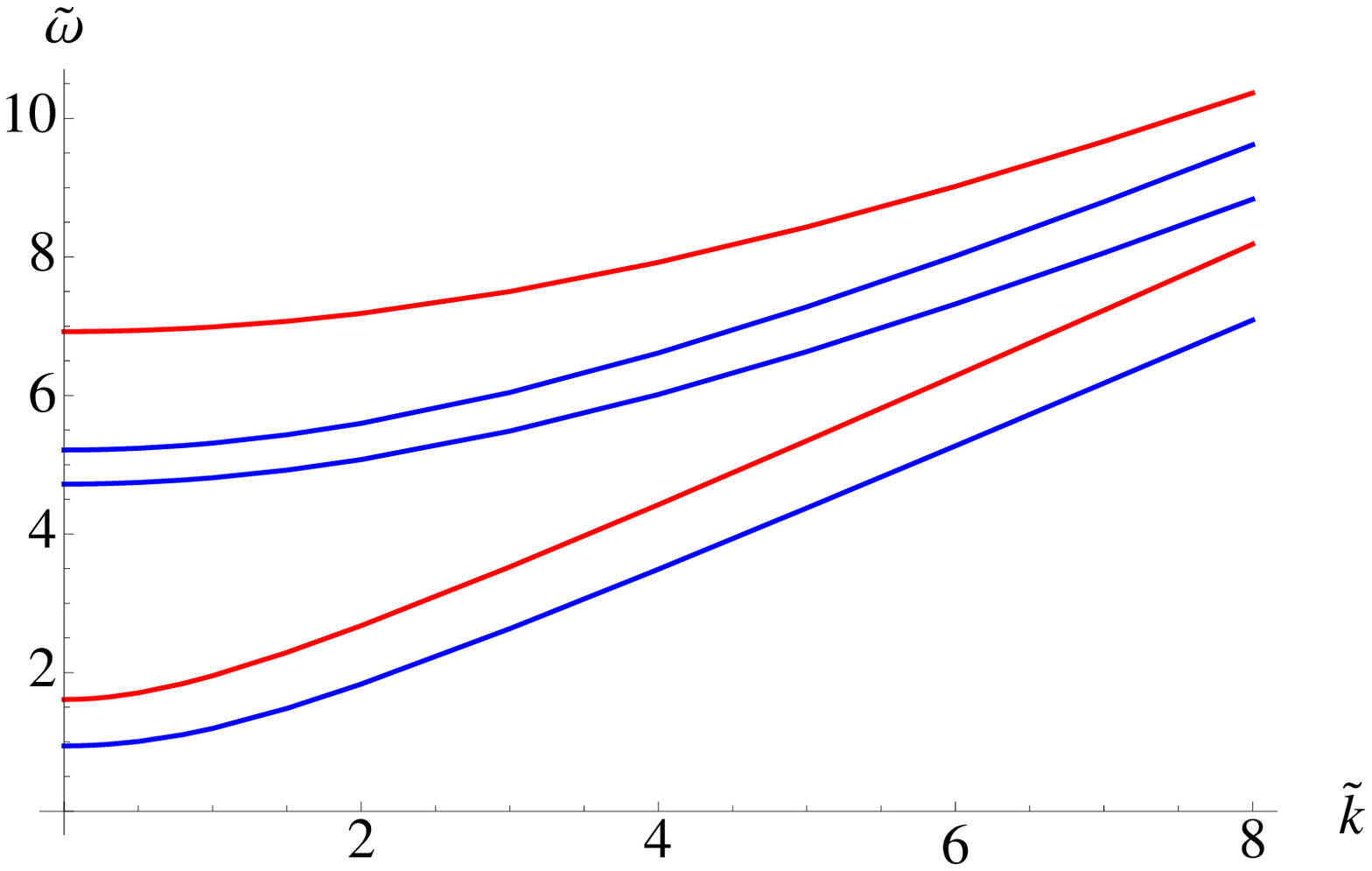}
\includegraphics[width=0.40\textwidth]{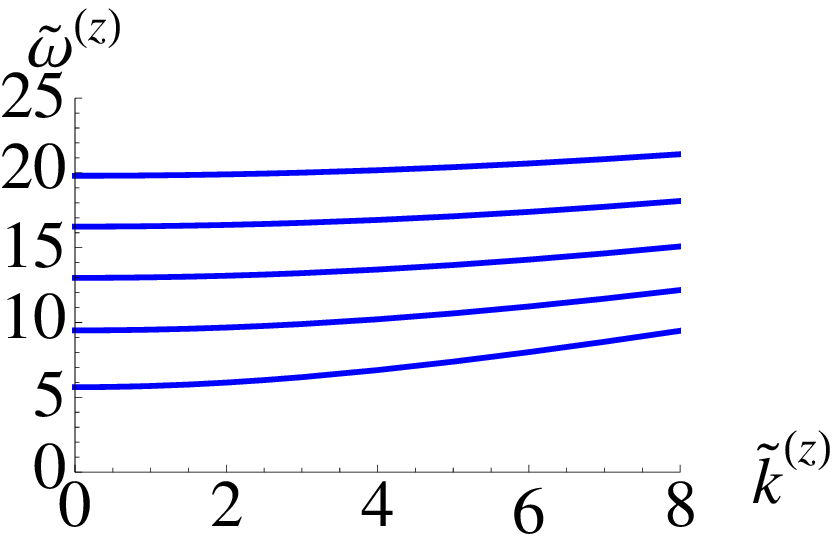}
\caption{In the left panel the dispersion relation of the vector fluctuations (blue) and of the scalar fluctuations (red, the first and the fourth curve from the top) 
at $\tilde m=-3$ and $\Delta_+=-1$. In the right panel the dispersion of $\delta z$ at $\tilde m=-3$ and $\Delta_+=-1$ is shown.}
\label{highdispersion}
\end{figure}

\begin{figure}[ht]
\center
\includegraphics[width=0.30\textwidth]{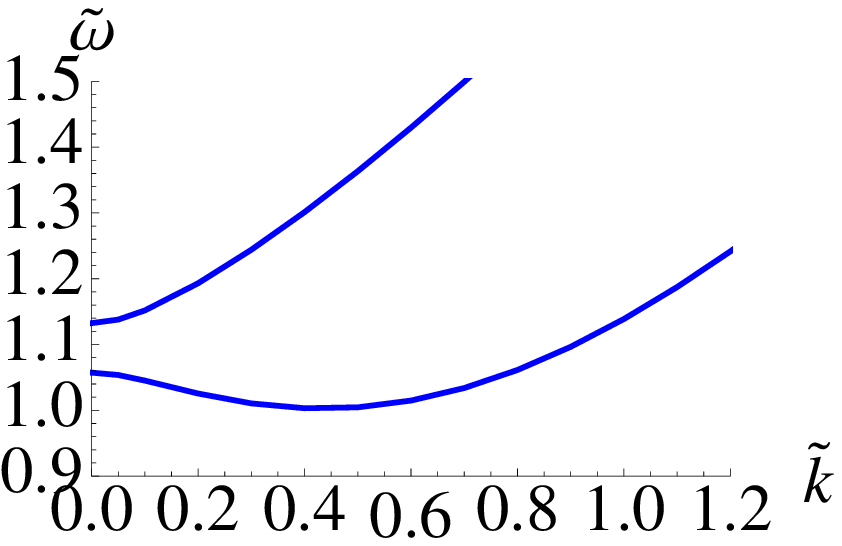}
\includegraphics[width=0.30\textwidth]{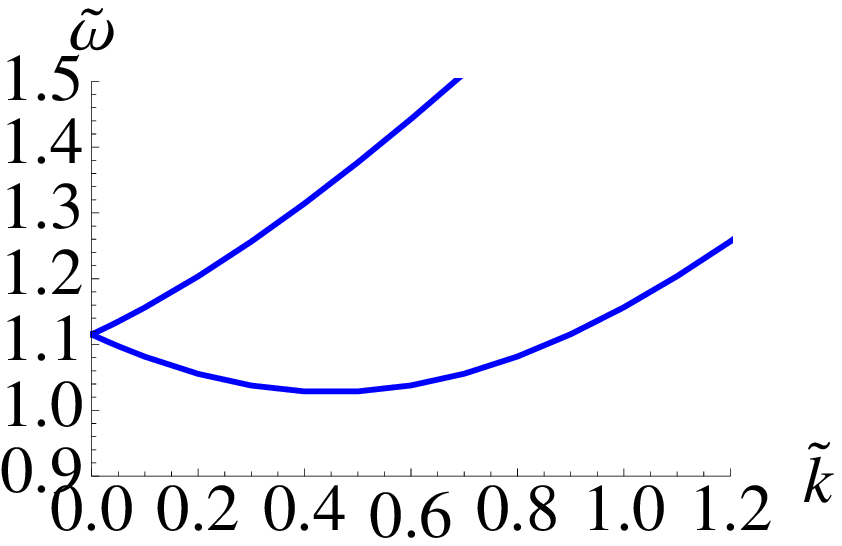}
\includegraphics[width=0.30\textwidth]{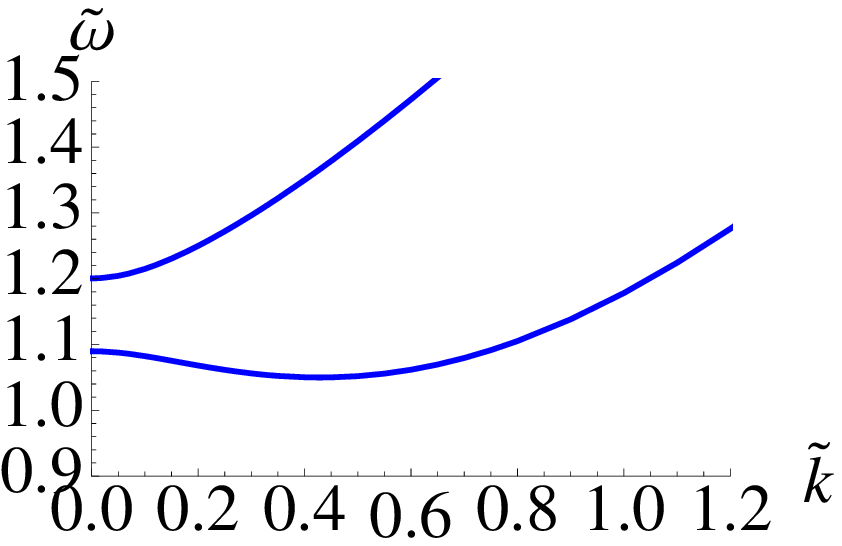}
\caption{The dispersion relation of the two lowest energy excitations for $\tilde m = -2.15, -2.185,$ and $-2.25$ from left to right, at $\Delta_{+}=-1$.  The lowest excitation displays a roton minimum $\tilde \omega= \tilde \omega_*$ at nonzero $\tilde k = \tilde k_*$.}\label{roton1}
\end{figure}

We now turn to the analysis of the dispersion relations of these excitations, starting with the higher mass fluctuations. 
Their dispersions are plotted in Fig.~\ref{highdispersion} and are of the form
\begin{equation}
 \tilde \omega=\sqrt{\tilde \omega_0^2+c^2_s\tilde k^2} \ ,
\end{equation}
where $c_s$ is the speed of sound for these fluctuations.  Interestingly, we find that for sufficiently large momentum some of the lines cross each other, 
which means that there is some degeneracy.  We also find that as the magnetic field is increased (for fixed filling fraction), $c_s$ increases towards one.

We now focus on the more interesting case of the two lowest energy modes, one of scalar origin and one of vector origin. 
As the magnetic field is increased (for fixed filling fraction), the lowest energy scalar mode decreases in energy, while the vector mode increases slightly.  As depicted in Fig.~\ref{roton1}, at some critical magnetic field the lowest energy mode becomes a roton; this is, it develops a minimum at nonzero momentum.  The dispersion assumes the characteristic rotonic form:
\begin{equation}\label{eq:fit}
 \tilde{\omega}=\sqrt{\tilde \omega_*^2+c_s^2(\tilde k-\tilde k_*)^2} \ ,
\end{equation}
where the star corresponds to the values at the lowest point of the dispersion curve for given $\tilde m$.   Although at this minimum the roton carries a nonzero momentum $\tilde k_*$, its group velocity is zero. 

\begin{figure}[ht]
\center
\includegraphics[width=0.40\textwidth]{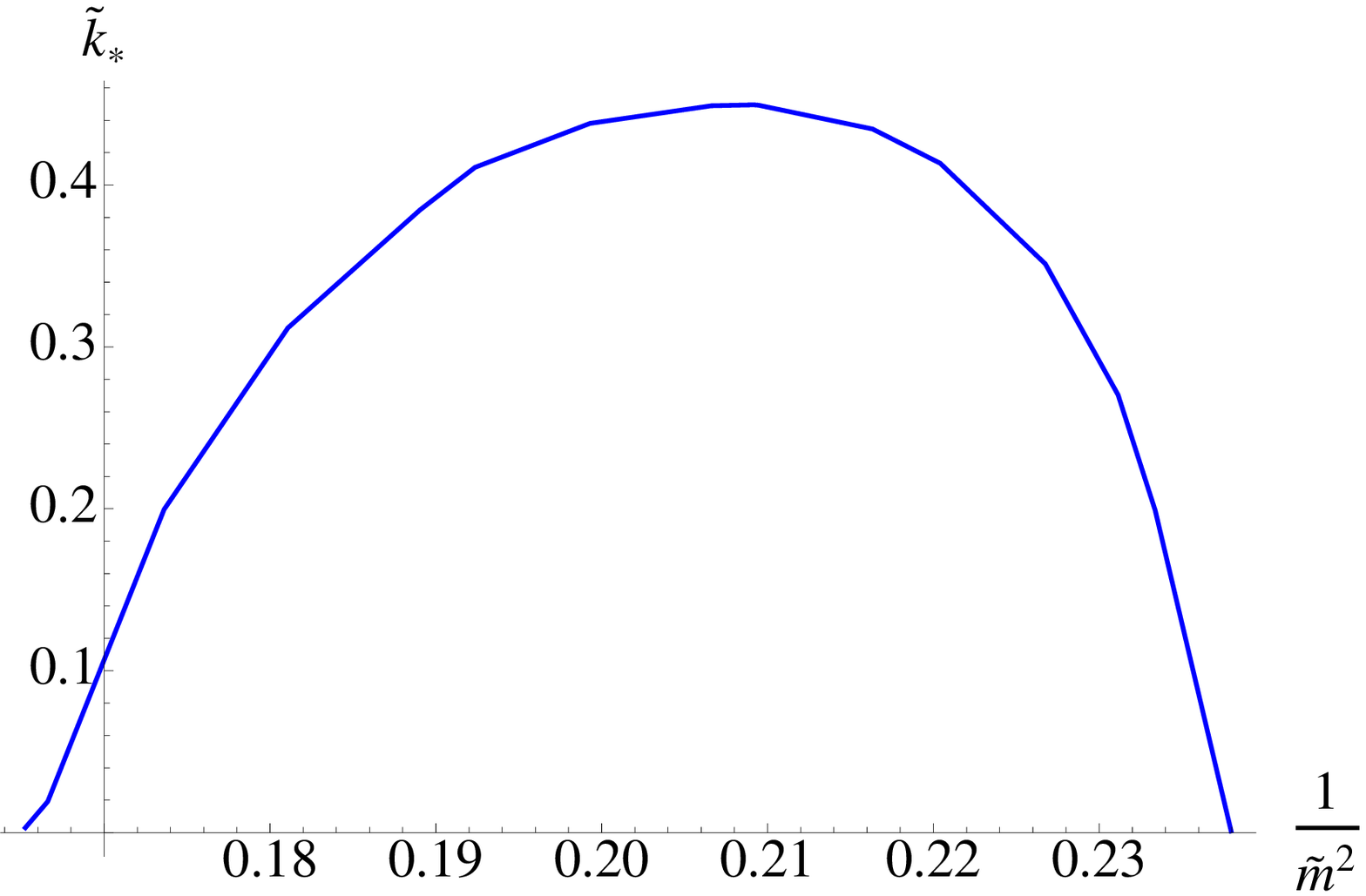}
\includegraphics[width=0.40\textwidth]{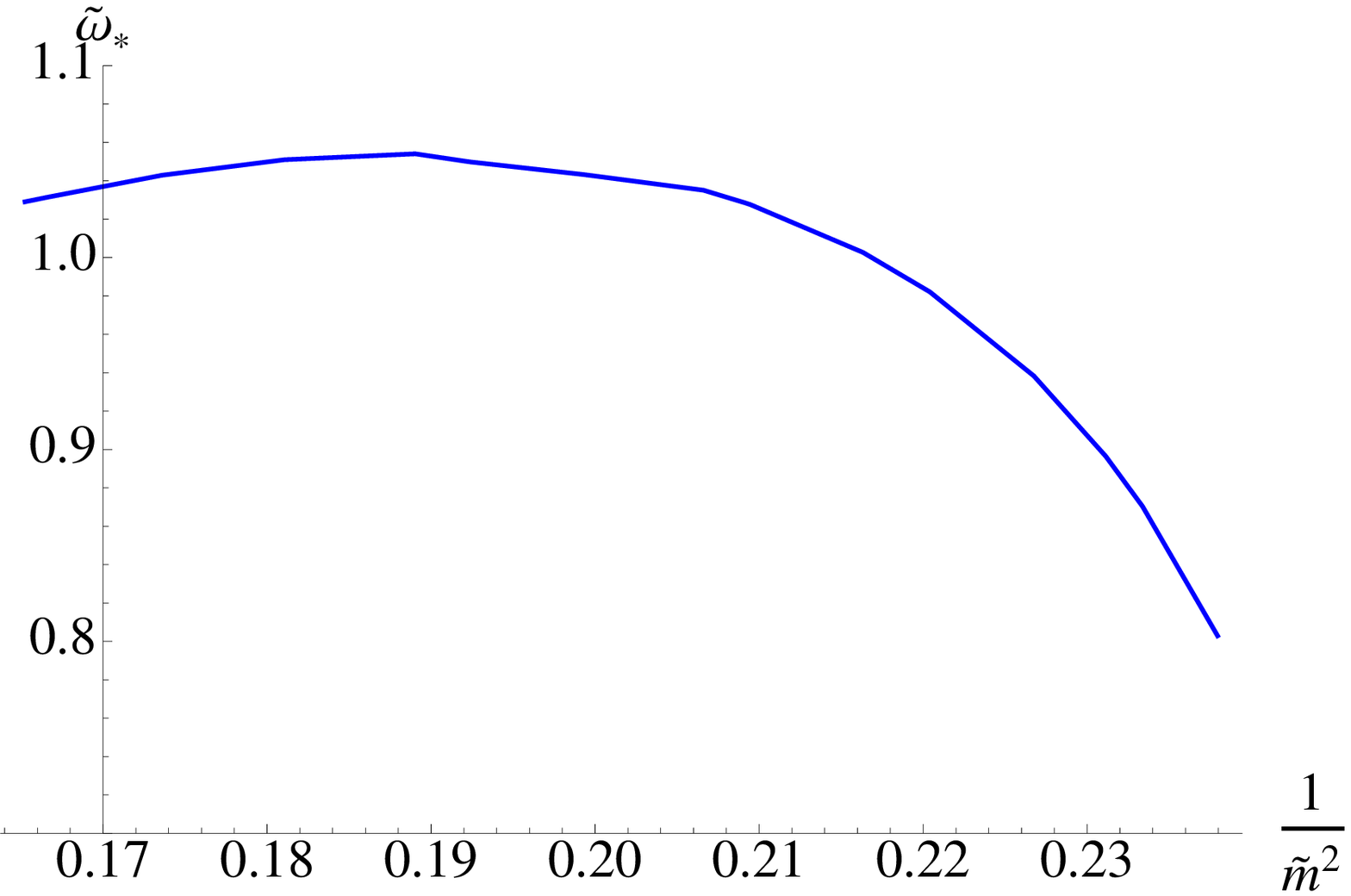}
\caption{The location $\tilde k_*$ of the roton minimum as a function of $1/\tilde{m}^{2}$ is shown in the left panel. The right panel displays
the minimum of the roton energy as a function of $1/\tilde{m}^{2}$. Both figures are for $\Delta_+=-1$.}
\label{roton2}
\end{figure}

As the magnetic field is increased further, the roton minimum $\tilde\omega_*$ initially gets deeper, then $\tilde k_*$ begins decreasing, until finally the minimum  
disappears as $\tilde k_*\to 0$. We depict this behavior in Fig.~\ref{roton2} in units of the magnetic length.
In this range of $\tilde m$ where the roton exists, it is always the lowest energy excitation of the system.
Notice that the roton only exists in the range of parameters where there is also a neighboring excitation. This suggests that the near degeneracy is
correlated with the existence of the roton.

This behavior is seen for all filling fractions, and the region of parameter space where the roton exists is around the region of 
parameter space where the energy of the two lowest excitations become close. The existence of two very close energy 
excitations has been recently observed in experiments \cite{pinchuz} and has received some theoretical explanation from 
hydrodynamics \cite{tokatly} and from the composite fermion picture \cite{jain1,jain2}.

\begin{figure}[ht]
\center
\includegraphics[width=0.30\textwidth]{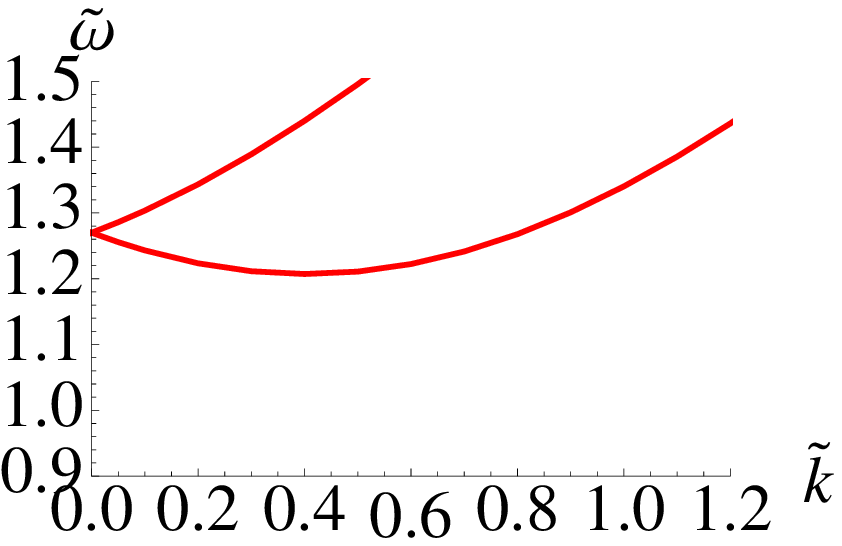}
\includegraphics[width=0.30\textwidth]{dispersion_tildemminus2185.eps}
\includegraphics[width=0.30\textwidth]{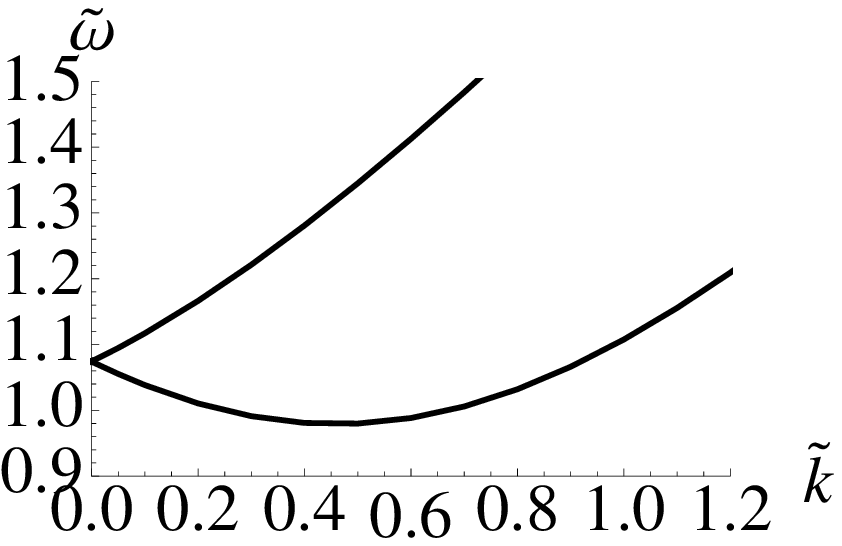}
\caption{Dispersion relations for the two lowest energy states when their energy is degenerate at zero momentum: (left) $\Delta_{+}=-1/2,\frac{\nu}{N}\approx 0.7426$,
(middle) $\Delta_+=-1,\frac{\nu}{N}\approx 0.7082$, and (right) $\Delta_{+}=-5/4,\frac{\nu}{N}\approx 0.6999$. Notice that the minimum energy $\tilde\omega_*$ decreases with decreasing 
filling fraction.}
\label{differentfilling}
\end{figure}

In general, we find that as the filling fraction decreases, the minimum energy of the excitation decreases as well.  
In addition, the value of  $\tilde m$ at which the roton appears also decreases with decreasing filling fraction.  To illustrate this, 
we plot in Fig.~\ref{differentfilling} the dispersion curves for various values of $\nu/N$, with $\tilde m$ chosen for each 
one such that the two lowest excitations are degenerate at zero $\tilde k$.  We see that $\tilde\omega_*$ goes down with 
decreasing $\nu/N$ but does not reach zero in the range of (\ref{eq:range}) of possible filling fractions.

\section{Discussion}

In this paper we continued to investigate the holographic description of a quantum Hall fluid first
described in \cite{Bergman:2010gm}. This D3-D7' model was shown previously to have the zero longitudinal conductivity 
and the correct Hall conductivity appropriate for a quantum Hall fluid.  Here we analyzed the neutral sector and showed 
that it is gapped as well, leading to an incompressible fluid.  We also showed that the lowest neutral excitation has a dip in its 
dispersion relation leading to its identification as a magneto-roton. 

Another intriguing similarity with observed quantum Hall systems is the existence of an excitation with ordinary dispersion slightly 
above the rotonic excitation (as shown in Fig.~\ref{roton1}). At small momentum, the non-rotonic excitation was observed in \cite{pinchuz} to 
be very close to the rotonic one.  In our model the two are close only in a range of densities; it is unclear in real materials what role the 
density plays in this picture. We also note that it is believed that the non-rotonic excitation is a two-roton bound state \cite{jain2}, 
while in the D3-D7' model, it is just one of the mesons.

While we found the model agrees with many expected properties of the quantum Hall fluid, there are also some significant differences.  
In terms of the fluctuation spectrum, although we have reproduced a magneto-roton, it does not completely match with those observed in 
experimental quantum Hall systems.

In this model there is a large hierarchy between the lowest energy neutral excitation, which is of the order of $\sqrt{B}(g_{YM}^{2}N)^{-1/4}$, and 
the energy scale of charged excitations, which is expected to be of the order of $\sqrt{B}(g_{YM}^{2}N)^{1/4}$. This is rather different from 
real quantum Hall systems where there is no such parametric gap.  Thus in the holographic model, where $g_{YM}^{2}N$ is large, the lowest neutral 
excitations are very deeply bound quasiparticle-quasihole bound states. This difference in energy is geometrically clear from the point of 
view of the bulk; quasiparticles are expected to be formed by long strings stretched from the tip of the D7-brane to the horizon (probably 
bound to some wrapped five branes), and the lowest neutral excitations are short strings representing fluctuations of the D7-brane. 

The magneto-roton we found seems to exist only for a range of magnetic fields (or densities), and we found just one minimum in the dispersion 
relation for any filling fraction. This is in contrast to real materials where the number of roton minima depends on the filling fraction 
and, as far as we know, is not sensitive to the density.

Although our results support the proposal of \cite{Bergman:2010gm} that the D3-D7' system holographically describes a quantum 
Hall fluid,  it is an imperfect model.  This is not entirely surprising, considering that holographic models necessarily have 
various unphysical properties, for example large $N$, which are undesirable from the point of view of applications to 
physical systems.  However, because the quantum Hall effect is a generic feature of (2+1)-dimensional fermion systems with broken parity 
and exhibits many universal properties, we are ultimately optimistic about our chances to describe it holographically.

\bigskip
\noindent

{\bf \large Acknowledgments}

We thank Oren Bergman, Matti J\"arvinen, Matthias Kaminski, Esko Keski-Vakkuri, Bom-Soo Kim, Rene Meyer, Sean Nowling, Efrat Shimshoni, Ady Stern, and Jan Zaanen for useful comments and discussions.  N.J. has been supported in part by the Israel Science Foundation under grant no.~392/09 and in part at the Technion by a fellowship from the Lady Davis Foundation.  
The work of G.L.  is supported in part by the Israel Science Foundation under grant no.~392/09. G.L wishes to thank the Erwin Schr\"odinger institute for hospitality.  
The research of M.L. is supported by the European Union grant FP7-REGPOT-2008-1-CreteHEPCosmo-228644 and in part by the APCTP Focus Program 
Aspects of Holography and Gauge/string duality at the Asia Pacific Center for Theoretical Physics (APCTP).  M.L. would like to thank 
the APCTP and the Galileo Galilei Institute for Theoretical Physics for their hospitality while this research was in progress.

\end{document}